\documentclass[american,aps,prb,superscriptaddress, twocolumn,
showpacs,amsmath,amssymb,floatfix]{revtex4}
\usepackage[latin9]{inputenc}
\usepackage{amsmath}
\usepackage{graphicx}
\usepackage{amssymb}
\usepackage{esint}
\usepackage{bm}
\usepackage{verbatim}
\usepackage{color}
\usepackage[normalem]{ulem}

\makeatletter
\@ifundefined{textcolor}{}
{%
 \definecolor{BLACK}{gray}{0}
 \definecolor{WHITE}{gray}{1}
 \definecolor{RED}{rgb}{1,0,0}
 \definecolor{GREEN}{rgb}{0,1,0}
 \definecolor{BLUE}{rgb}{0,0,1}
 \definecolor{CYAN}{cmyk}{1,0,0,0}
 \definecolor{MAGENTA}{cmyk}{0,1,0,0}
 \definecolor{YELLOW}{cmyk}{0,0,1,0}
 }

\makeatother

\usepackage{babel}

\begin{document}


\title{Noise studies of magnetization dynamics in dilute magnetic semiconductor heterostructures}

\author{Vikram Tripathi}
\affiliation{Department of Theoretical Physics, Tata Institute of Fundamental Research, Homi Bhabha Road, Navy Nagar, Mumbai 400005, India}

\author{Kusum Dhochak}
\affiliation{Department of Theoretical Physics, Tata Institute of Fundamental Research, Homi Bhabha Road, Navy Nagar, Mumbai 400005, India}

\author{B.A.~Aronzon}
\affiliation{National Research Center ``Kurchatov Institute'',
Kurchatov Square 1, Moscow, 123182 Russia}
\affiliation{Institute for Theoretical and Applied Electrodynamics, Russian Academy of Sciences, Izhorskaya Str. 13, Moscow, 125412 Russia}

\author{Bertrand Raquet}
\affiliation{Laboratoire National des Champs Magn\'{e}tiques Intenses, Universit\'{e} de Toulouse, UPS, INSA, CNRS-UPR 3228, 143 avenue de Rangueil, F-31400 Toulouse, France}

\author{V.V.~Tugushev}
\affiliation{National Research Center ``Kurchatov Institute'',
Kurchatov Square 1, Moscow, 123182 Russia}

\author{K.I.~Kugel}
\affiliation{Institute for Theoretical and Applied Electrodynamics, Russian Academy of Sciences, Izhorskaya Str. 13, Moscow, 125412 Russia}

\begin{abstract}
We study theoretically and experimentally the frequency and temperature dependence of resistivity noise in semiconductor heterostructures $\delta$-doped by Mn. The resistivity noise is observed to be non-monotonous as a function of frequency. As a function of temperature, the noise increases by two orders of magnitude for a resistivity increase of about 50\%. We study two possible sources of resistivity noise -- dynamic spin fluctuations and charge fluctuations, and find that dynamic spin fluctuations are more relevant for the observed noise data. The frequency and temperature dependence of resistivity noise provide important information on the nature of the magnetic interactions. In particular, we show how noise measurements can help resolve a long standing debate on whether the Mn-doped GaAs is a $p$-$d$ Zener/Ruderman-Kittel-Kasuya-Yosida (RKKY) or double exchange ferromagnet. Our analysis includes the effect of
different kinds of disorder such as spin-glass type of interactions
and a site-dilution type of disorder. We find that the resistivity
noise in these structures is well described by a disordered RKKY ferromagnet model dynamics with a conserved order parameter.
\end{abstract}


\date{\today}

\maketitle

\section{Introduction \label{sec:Introduction}}

The mechanism of magnetic ordering in dilute magnetic semiconductors (DMSs) has traditionally attracted much attention of investigators. One of the issues being most intensely studied~\cite{matsukura98,ohno01,dietl01,dietl07,awschalom02,sandratskii02, zutic04,priour04,mahadevan04,jungwirth06,sato10,furdyna88} concerns the microscopic nature of magnetism in bulk DMSs; in particular, whether the microscopic interaction governing the ferromagnetism (FM) is of Zener/RKKY type or double-exchange type. Let us recall that the Zener/RKKY model is applicable when the effective coupling between local magnetic moments and the spins of charge carriers is smaller than the carrier bandwidth, and, this model is widely used in the context of magnetism in the Kondo lattice systems~\cite{doniach77,newns87,broholm87,schroder00}. On the other hand, the double exchange model is believed to describe the magnetism in manganites~\cite{salamon01} and double perovskites~\cite{kobayashi98} where the Hund's rule coupling is known to be large compared to the carrier bandwidth. While the ordering of local moments of magnetic metal in both cases is mediated by the charge carriers of the semiconductor (SC) host, the two models differ in the relative importance of the magnitudes of the effective coupling of the local moments with the spins of charge carriers and the intersite hopping integrals (in other words the bandwidth for charge carriers). The current opinion is that both mechanisms of FM can take place in bulk DMSs. The conditions of the realization of either mechanism strongly depend not only on the concentration of magnetic metal ions, but also on the character of their distribution in the SC host. The last factor plays an important role in FM ordering in bulk DMSs and is crucially driven by the details of their growth technology. Besides, technological or fabrication aspects drastically affect the character of fluctuations of crystal-field and exchange potentials in the SC host as well as the degree of ionization of magnetic metal atoms and effective concentration of carriers supplied by them when doping the host. These aspects also determine the relative importance of phase segregation in the system, as well as the contribution of wide and narrow bands in the spectrum of carriers to the coupling between local magnetic moments of the metal.

An emerging trend in the studies of DMSs concerns the investigation of hybrid layered heterostructures (in the following we shall also use the term ``2D DMS structures'') containing magnetic ultrathin metal layers (so-called $\delta$-layers) embedded into a nonmagnetic SC heterostructure. This interest in 2D DMS structures has a twofold motivation. The first one is practical: present-day microelectronic devices have a planar geometry and it becomes necessary to understand the mutual interaction of different parts of the structure such as the $\delta$-layer and quantum wells. A number of studies of 2D DMS structures have been reported in the literature~\cite{awschalom04,nazmul05,wojtowicz03,aronzon10,
aronzon07,vasilieva05,wurstbauer08,rupprecht10,dietl10,dietl10a}. In Refs.~\onlinecite{nazmul05,wojtowicz03}, the FM state in Mn $\delta$-doped GaAs/AlGaAs heterostructures was observed at rather high temperatures. However, these devices have low mobility values because the authors of Refs.~\onlinecite{nazmul05,wojtowicz03} aimed at achieving the highest possible hole density just in the vicinity of Mn ions to maximize the Curie temperature $T_{C}$. Note here that Mn ions are responsible not only for magnetism but also act as acceptors and a source of the random potential that limits mobility. Second, properties of 2D magnets are often qualitatively different from those of their bulk counterparts, and studies in heterostructures can sometimes clarify issues of basic research on two-dimensional magnetism in solids. Issues peculiar to the heterostructures, such as magnetic anisotropy, limit their use in a general 2D ferromagnet problem. However for properties such as the frequency dependence of the magnetization noise, the universality of the behavior is not sensitive to issues such as magnetic anisotropy since the universality is dictated by other properties such as the existence of a conserved order parameter, and heterostructures provide a good platform. Note also that as we have studied resistivity noise of the hole gas in the 2D quantum well, our analysis will be useful for studies of magnetic impurity effects in low-dimensional conductors.  It is also perhaps worth observing here that the idea of using resistivity noise for probing magnetization dynamics that we develop here is not specifically for heterostructures and can be adapted for the study of other 2D ferromagnets.

As a rule, in the bulk DMSs there are no insurmountable problems in obtaining detailed information on FM ordering from direct magnetic measurements although it is often difficult to correctly interpret the obtained data in terms of some specific theoretical model. In 2D DMS structures, the situation turns out to be more complicated and direct magnetic measurements that are able to detect FM ordering in the metal layers are often tedious, troublesome, and sometimes even spurious as opposed to the bulk DMSs. Moreover, direct measurements of the magnetization are impractical given the small size of the magnetically active region in 2D DMS structures. In such a situation, an indirect retrieval of magnetic characteristics (for example, from resistivity and Hall effect measurements) becomes decisive, whereas for the bulk DMSs, it is in general a subsidiary tool. However, even in the bulk DMSs, the resistivity anomaly, which is a peak or shoulder at the temperature dependence of resistivity, is widely used as an evidence for the onset of significant FM correlations and the measure of $T_{C}$.~\cite{jungwirth06}  In an earlier paper (Ref.~\onlinecite{tripathi11}), it was demonstrated that in contrast to the resistivity anomaly in bulk DMSs, which typically appears near (and above) the Curie temperature, the resistivity anomaly observed in 2D DMS structures typically appear far below the mean-field Curie temperature and may even exist in the absence of an actual magnetic phase transition. Nevertheless, the resistivity anomaly (a peak or shoulder) in 2D DMSs can be also regarded as an indication of the onset of significant FM correlations).~\cite{nazmul05,tripathi11}

The use of indirect experimental methods to reveal magnetic properties in 2D DMS structures introduces additional considerations into the interpretation of data compared to bulk samples. For example, magnetic proximity effects and long-range Coulomb interaction between different components of such structures have a crucial influence on the magnetic and transport characteristics of the system. Recently, structures with a single FM $\delta$-layer deposited near or inside the SC quantum well (QW) forming the 2D channel of conductivity have emerged as a test system to study magnetic proximity effects. For example, in Ref.~\onlinecite{awschalom04}, photoluminescence polarization was observed in  Al$_{0.4}$Ga$_{0.6}$As/GaAs/Al$_{0.4}$Ga$_{0.6}$As structures with a 0.5 monolayer of MnAs placed at 5 nm from a GaAs QW. Crucially, these structures had a gate, which allowed shifting holes from the QW to the FM layer. It was shown that the holes in the QW are spin-polarized and that their degree of polarization strongly increases (from 0.4\% to 6.3\%) with the ``pressing'' gate voltage. In the control structures, lacking the FM $\delta$-layer, the photoluminescence polarization was not observed. Induced photoluminescence polarization effect was recently observed in other structures such as GaAs/In$_x$Ga$_{1-x}$As/GaAs ($x \approx 0.2$) containing Mn $\delta$-layer separated from the InGaAs QW by a 3-5 nm thick GaAs spacer.~\cite{zaitsev09} It is clear that the aforementioned results can hardly be explained by a simple tunneling of holes from the QW to FM layer through the spacer, since even under a pressing voltage the characteristic depth of such tunneling does not exceed 1 nm.

In 2D DMS structures containing spatially separated 2D hole gas in the QW and FM $\delta$-layers, the mutual influence of these two subsystems has three effects that have a bearing on the magnetic properties. First, interpenetration of the wave function tails between the QW and the FM layers polarizes the spins of itinerant charge carriers in the QW and modifies the effective coupling between local spins in the FM layer.~\cite{awschalom04,meilikh08-11} Second, quantum fluctuations in the QW stabilize magnetic order in the FM layer, suppressing at the same time the amplitude of the magnetic moment and the transition temperature with respect to those found by mean-field estimates.~\cite{melko07} Third, electrostatic charge redistribution occurs between the QW and the FM layers due to their different density of states and depths; and as a result, modification of the magnetic characteristics of the FM layer occurs even on purely classical grounds.~\cite{caprara11}

Recently, different ways have been proposed to describe self-consistently ferromagnetic ordering and spin polarization of charge carriers in the 2D DMS structures under discussion. Assuming that Mn atoms form isolated paramagnetic centers inside the $\delta$-layer, the authors of Refs.~\onlinecite{meilikh08-11} proposed a model of indirect RKKY-type exchange coupling between local moments of Mn atoms in the $\delta$-layer via the "tails" of the wave function of itinerant hole states in QW. In the framework of the same assumption about of a localized character of the hole states in the $\delta$-layer, the authors of
Ref.~\onlinecite{averkiev12} formulated the problem in terms of the Anderson-Fano model of configuration interaction between the localized hole states at Mn centers and itinerant hole states in the QW. In the framework of the method of Ref.~\onlinecite{averkiev12},  the spin splitting of itinerant hole states in an external magnetic field is strongly enhanced due to their hybridization with localized hole states. This leads to a resonant enhancement of the interband radiation recombination of itinerant holes and causes circular polarization of the luminescence in QW.

On the other hand, an alternative approach based on the assumption of an itinerant character of the hole states in both the QW and  FM $\delta$-layer was successfully used~\cite{menshov09,caprara09}. It has been shown that a thin layer of FM metal located in a bulk SC matrix induces quasi-2D spin-polarized collective states and a half-metallic type of an electron spectrum of the system is formed~\cite{menshov09,caprara09}. These states (called also confinement states) have rather extended character along the structure axis of growth because of their small binding energies, so they penetrate deep into the spacer and can reach the QW. It may be shown (see the next Section) that hybridization of confinement states in the FM $\delta$-layer, with one-electron states in QW leads to the spin polarization of carriers in the QW.

In the model~\cite{menshov09,caprara09}, the holes in the QW play mostly an ``observer'' (passive) role in the exchange coupling inside the $\delta$-layer, although this does not mean that they have no effect on the FM ordering in the  system.~\cite{meilikh08-11,averkiev12} If one assumes an easy-plane character of magnetic anisotropy of the ``free'' $\delta$-layer (i.e. in the absence of the QW), then introducing a QW near the $\delta$-layer can drastically change the type of magnetic anisotropy. As a result, the easy axis of magnetization appears to be directed along the normal to the $\delta$-layer plane.~\cite{korenev03} Experiments~\cite{awschalom04,zaitsev09,aronzon10} as well as recent direct measurements of magnetic anisotropy~\cite{rupprecht10} give clear indication of the existence of such an orientational  transition.

In addition to the effects of induced spin polarization of holes in the QW and the reorientation of magnetization in the FM $\delta$-layer, there exist purely electrostatic (Coulomb) effects of charge redistribution between these two subsystems. As it was shown in Ref.~\onlinecite{tripathi11}, the long-range fluctuations of electrostatic potential, which are inevitable due to an inhomogeneous distribution of magnetic metal ions over the FM $\delta$-layer, are projected onto the QW and give rise to similar fluctuations of electrostatic potential in the QW. Due to this fact, the transport measurements involving charge carriers in the QW can reveal some tiny details of disorder in the distribution of magnetic metal ions in the FM $\delta$-layer, the latter being inaccessible by the direct magnetic measurements.

It is evident from the above discussion that in comparison with bulk DMS systems, 2D DMS structures have a number of additional experimental and theoretical complications arising from the use of indirect probes for magnetism as well as the interplay between the magnetic $\delta$-layer and the QW. The choice of the indirect probe is dictated not only by the convenience of measurement, but also by the level of detail one is seeking as regards the properties.

Resistivity measurement, although an indirect probe of magnetism, is one of the most widely used probes in both bulk and 2D DMS systems given its simplicity. In both systems, the observation of an anomaly in the temperature dependence of the resistivity is associated with the onset of significant ferromagnetic correlations. Unfortunately, such a measurement is unsuitable for shedding light on the microscopic interactions governing FM: both the RKKY and double-exchange models can describe the resistivity anomaly observed in DMSs. This is not surprising since as far as such static properties are concerned, both can be described by effective Heisenberg models. The differences manifest themselves, however, in the dynamic properties such as the autocorrelation function, since the damping of magnetic excitations is sensitive to the microscopic details governing the fluctuations. In the following, we shall show how the magnetization dynamics can be probed through resistivity noise measurements. While our analysis is applied to 2D DMS structures, we note that this method of deducing the mechanism of FM from the analysis of dynamic susceptibility is quite general and can be used for other systems.

\begin{figure}
\begin{centering}
\includegraphics[width=0.65\columnwidth]{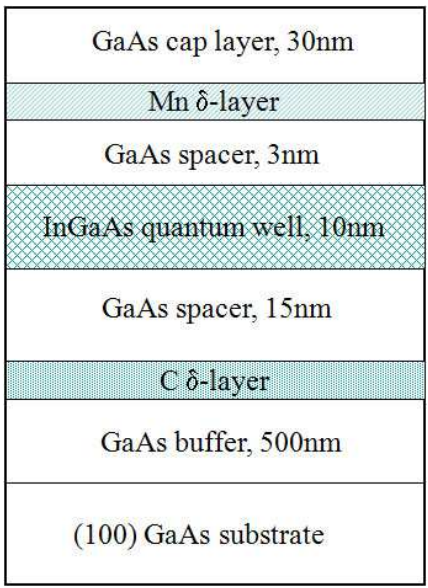}
\par\end{centering}
\caption{\label{fig:structure} (Color online) Layout of the Mn $\delta$-doped GaAs/In$_{x}$Ga$_{1-x}$As/GaAs QW heterostructures used in this work and in Ref.~\onlinecite{tripathi11}. The carbon (C) layer is a nonmagnetic source of holes. Noise measurements reported in this paper correspond to Sample 4 in Ref.~\onlinecite{tripathi11}, an insulating sample on the border of a percolative metal-insulator transition.}
\end{figure}

In this paper, we present a theoretical and experimental study of resistivity noise as an indirect probe of magnetic dynamics in 2D DMS structures $\delta$-doped by magnetic atoms. We show how resistivity noise can distinguish between RKKY/Zener and double-exchange mechanisms. We find that in our structures the noise measurements are consistent with a disordered RKKY picture. The experiments reported in this paper were performed using the GaAs/In$_{x}$Ga$_{1-x}$As/GaAs QW structure $\delta$-doped by Mn. Such structures produced by selective doping exhibit a high enough hole mobility (more than 2000 ${\rm cm}^{2}/{\rm V\cdot s}$ at 5 K)~ and show clear evidence of a 2D excitation spectrum as well as FM at moderately high temperatures of about $30{\rm K}$ \cite {aronzon10,aronzon07}. In Fig.~\ref{fig:structure}, we illustrate a schematic layout of the fabricated structures. Similar 2D heterostructures were also reported elsewhere; however, the ferromagnetic ordering was obtained at a much lower (millikelvin) range of temperatures.~\cite{wurstbauer08,rupprecht10,dietl10a}.

Depending on Mn content, the  samples of the type presented in Fig.~\ref{fig:structure} exhibit metallic or activation conductivity on both sides of the metal-insulator transition. The insulating samples are most interesting as they can give us valuable insights into the mechanism of ferromagnetism in these DMS heterostructures and we will focus on the sample very close to the percolation transition since it demonstrates some features of both metallic and insulating behavior.

The rest of the paper is organized as follows. In Sec.~\ref{sec:Proximity}, we clarify the role of magnetic proximity effects in the interplay of ferromagnetism and carrier transport in 2D DMS structures with a spatially separated FM $\delta$-layer and the QW. In Sec.~\ref{sec:Models}, we present our model for the
resistivity noise in QW and obtain its relation to charge and magnetization dynamics in the FM $\delta$-layer. It is shown how magnetization dynamics is sensitive to the microscopic nature of FM correlations and can be used to distinguish between RKKY and double-exchange types of coupling. Disorder effects are also analyzed. In Sec.~\ref{sec:Comparison-with-experiment}, we compare these findings with our experimental data on frequency and temperature dependence of resistivity noise. It is argued that the data supports the picture of a disordered RKKY type of FM over a double-exchange type of FM. Section~\ref{sec:Discussion} contains a discussion of our results.

\section{Magnetic proximity effect in 2D DMS structures with remote FM $\delta$-layer and quantum well} \label{sec:Proximity}

Before proceeding to the point at issue of our work, in this section we shall try to shed additional light on a key physical problem of the systems under consideration, namely, how magnetic correlations in the FM $\delta$-layer affect charge transport in the QW. Let us recall the microscopic Anderson-type model for a single FM $\delta$-layer embedded into the SC host proposed in Refs.~\onlinecite{menshov09,caprara09}. It has been shown that FM order in such system can be attributed to the intrinsic physical properties of the $\delta$-layer. The relevant effects described by this model are the hybridization of the electron states of the metal and semiconductor atoms, the strong charge and spin redistributions around the $\delta$-layer, and the electron-electron correlation on the metal atom. Following Refs.~\onlinecite{menshov09,caprara09}, confinement states in the form of quasi-2D spin-polarized subbands  located within the SC band gap arise in the host near the $\delta$-layer. This leads to the half-metallic character of the electron spectrum of the system. Under certain approximations, a picture of the low energy branches of these confinement states can be qualitatively reproduced by a simple phenomenological Hamiltonian $H_{\delta}:$
\begin{align}
H_{\delta} & = H_{host} + V_{\delta}(z).
\label{eq:Hdelta}
\end{align}
Here the 3D Hamiltonian $H_{host}$  describes itinerant electron states in a bulk SC, $V_{\delta}(z)$ is the effective one-dimensional spin polarized potential of the  charge carriers in a single FM $\delta$-layer
 \begin{align}
V_{\delta}(z) & = -[V_1 + V_2 \bm{e} \cdot \bm{\sigma}]\delta(z+L) <0, \, \, (V_1, V_2) > 0.
\label{eq:Vdelta}
\end{align}

The $z$ axis is oriented along the direction of the structure growth, the FM $\delta$-layer lies in the $z=-L$ plane,  $\delta(z)$ is the delta-function, $V_1$ and $V_2$ are parameters derived from the model described in Ref.~\onlinecite{menshov09}, $\bm{\sigma}$ is the vector composed of the Pauli matrices in the spin space, $\bm{e}$  is the unit vector along  the direction of magnetization of the FM $\delta$-layer.

Let us introduce the energy parameters $\omega_{\alpha}(\bm{q}) < 0$ specifying the positions of the spin-polarized  subbands of confinement states described by Hamiltonian~\eqref{eq:Hdelta}  for a single FM $\delta$-layer in the absence of the QW, with the wave functions $\phi_{\alpha}(\bm{q}, z)$  and characteristic lengths $\lambda_{\alpha}(\bm{q})=1/\sqrt{2m}|\omega_{\alpha}(\bm{q})|$ ($\mathbf{q}$ is the crystal momentum in the ($x,y$) plane
perpendicular to the $z$ axis, $\alpha  = \pm$ is the spin-projection index in the diagonal representation for the ${\bf{\sigma}}$ matrix. Below, we discuss a half-metallic regime, when  $|\omega_{-}(\bm{q})|< |\omega_{+}(\bm{q})|$ and the $\phi_{-}(\bm{q}, z)$ states are empty at all $\mathbf{q}$, while the $\phi_{+}(\bm{q}, z)$ states are partially occupied depending on the Fermi level position.

The low-energy branches in the spectrum of charge carriers in a single nonmagnetic QW are successfully modeled by a phenomenological Hamiltonian $H_{QW}$
\begin{align}
H_{QW} & = H_{host} + V_{QW}(z).
\label{eq:H-QW}
\end{align}
Here $V_{QW}(z)$ is an effective one-dimensional potential of the charge carriers in a single nonmagnetic QW:
\begin{align}
V_{QW}(z) & = -U(z) < 0.
\label{eq:V-QW}
\end{align}

Without the effects of charge redistribution between the FM $\delta$-layer and QW one can simply assume that $U = U_0 > 0$ at $0 < z <W$ and  $U = 0$ at $z < 0$ and $z > W$, where $W$ is the nominal QW thickness; the charge redistribution leads to a more intricate form of $U(z)$ due to appearance of a QW/spacer interface.

Let us introduce the energy parameters $\varepsilon_{n\alpha}(\bm {q}) < 0$ specifying the position of the $n$th spin-degenerate subband of 2D states in the spectrum of QW described by Hamiltonian~\eqref{eq:H-QW} of a single QW in the absence of the FM $\delta$-layer ($n=0, 1,\ldots$). The corresponding wave functions $\psi_{n \alpha}(\bm{q}, z)$ exponentially decay with $|z|$ away from  QW, $|z| \gg l_{n\alpha}(\bm {q})$, where $l_{n\alpha}(\bm {q})=1/\sqrt{2m}|\varepsilon_{n\alpha}|$ are characteristic lengths, and $m$ is the effective mass of charge carriers. In the simplest case, only one spin-degenerate subband of 2D states, $\psi_{0 \alpha}(\bm{q}, z)$ with the parameters $\varepsilon_{0 \alpha}(\bm {q}) = \varepsilon_0 (\bm {q})$  and $l_{0 \alpha}(\bm {q}) = l_0(\bm {q}),$ exists in the QW.  The states $\psi_{0 \alpha}(\bm{q}, z)$ are partially occupied depending on the Fermi level position.

Let us now treat the 2D DMS structure containing both the FM $\delta$-layer and the QW as a triple layer quantum system with an effective one-dimensional potential $V(z)$ composed of two wells with different widths and depths separated by the barrier (spacer) of width $L$. We can write the model Hamiltonian $H_{struct}$ describing the low-energy states of the charge carriers in the following form:
\begin{align}
H_{struct} & = H_{host} + V(z).
\label{eq:Hstr}
\end{align}

Evidently, the effective potential $V(z)$ cannot be correctly modeled as a simple sum of the potentials $V_{QW}(z)$  and $V_{\delta}(z)$, due to a strong redistribution of the carriers between the QW and FM $\delta$-layer. In principle, using an appropriate form of $V(z)$,  one can calculate (analytically or numerically) the eigenenergies and  eigenstates of Hamiltonian~\eqref{eq:Hstr} at all values of the spacer thickness $L$, but this method seems to be tedious. However, to reveal the main physics of the magnetic proximity effect between the FM $\delta$-layer and QW in the quantum structure described by Hamiltonian~\eqref{eq:Hstr}, we consider  only the situation of a relatively thick spacer, $L \gg \max\{l_0(\bm {q}), \lambda_{\alpha}(\bm{q})\}$.  We assume that the minima of $\varepsilon_0(\bm {q})$ and $\omega_{\alpha}(\bm {q})$ lie at $\bm {q} = 0$ and the occupation number of $\phi_{+}(\bm{q}, z)$ states in FM $\delta$-layer significantly exceeds that of $\psi_{0 \alpha}(\bm{q}, z)$ states in the QW. Thus, in the half-metallic regime we have $|\omega_{+}(\bm {q})| > |\varepsilon_0(\bm {q})| > |\omega_{-}(\bm {q})|$ and $\lambda_{-}(\bm {q}) > l_0(\bm {q}) > \lambda_{+}(\bm {q})$ and the Fermi level $\mu < 0$ lies between two minima, $|\omega_{-}(0)|$ and $|\varepsilon_0(0)|$, i.e.  $|\omega_{+}(0)| > |\varepsilon_0(0)| > |\mu| > |\omega_{-}(0)|$.
This corresponds to the fact that the charge carrier density in the FM $\delta$-layer is much higher than that in the QW.

We now write the wave function $\chi_{\alpha}(\bm{q}, z)$ of the system with Hamiltonian~\eqref{eq:Hstr} in the variational form
\begin{align}
\chi_{\alpha}(\bm{q}, z) & = A_{\alpha}(\bm{q})\psi_{0 \alpha}(\bm{q}, z) + B_{\alpha}(\bm{q})\phi_{\alpha}(\bm{q}, z),
\label{eq:chi}
\end{align}
where $A_{\alpha}(\bm{q})$ and $B_{\alpha}(\bm{q})$ are coefficients obeying an evident normalization condition:
\begin{align}
|A_{\alpha}(\bm{q})|^2 + |B_{\alpha}(\bm{q})|^2 + 2A_{\alpha}(\bm{q})B_{\alpha}(\bm{q})S_{\alpha}(\bm{q}) & = 1,
\label{eq:AB-norm}
\end{align}

\begin{align}
S_{\alpha}(\bm{q}) & = \int \psi_{0 \alpha}(\bm{q}, z)\phi_{\alpha}(\bm{q}, z) dz.
\label{eq:S-alpha}
\end{align}
Note that the functions $\psi_{0 \alpha}(\bm{q}, z)$ and $\phi_{\alpha}(\bm{q}, z)$ are not orthogonal, so $S_{\alpha}(\bm{q}) \neq 0$.

Obviously, this simplest approach treats only qualitatively the problem of interference of the charge carrier states of two subsystems as a simple superposition of the wave functions $\psi_{0 \alpha}(\bm{q}, z)$ and $\phi_{\alpha}(\bm{q}, z)$.  Moreover, it is unable to capture in full measure possible resonance effects (for example, arising additional bound states). However, even with such an approximate method, we can project the states $\phi_{\alpha}(\bm{q}, z)$ of the FM $\delta$-layer onto the states $\psi_{0 \alpha}(\bm{q}, z)$ of QW using a redefinition $V(z)= \tilde{V}_{\delta}(z) + V_{QW}(z)$, where $\tilde{V}_{\delta}(z) = V(z) - V_{QW}(z)$  and treating  $\tilde{V}_{\delta}(z)$ as a small perturbation to $V_{QW}(z)$ at  $L \gg \max\{l_0(\bm {q}), \lambda_{\alpha}(\bm{q})\}$. Omitting the cumbersome but straightforward calculations, we write the final expression for the energy spectrum of the charge carriers in QW in the second order of the perturbation series expansion in terms of $\tilde{V}_{\delta}(z)$:
\begin{align}
E_{\alpha}^{QW}(\bm{q}) & = \varepsilon_0(\bm{q}) + \Delta_{\alpha}^{QW}(\bm{q}),
\label{eq:E-alpha}
\end{align}
where
\begin{align}
\Delta_{\alpha}^{QW}(\bm{q}) & \approx  \tilde{V}_{\delta}^{' \alpha}(\bm{q}) +[\tilde{V}_{\delta}^{'' \alpha}(\bm{q})]^2 /[\varepsilon_0(\bm{q}) - \omega_{\alpha}(\bm {q})],
\label{eq:Delta-alpha}
\end{align}

\begin{align}
\tilde{V}_{\delta}^{' \alpha}(\bm{q}) & =\int \psi_{0 \alpha}(\bm{q}, z)\tilde{V}_{\delta}^{\alpha}(\bm{q})\psi_{0 \alpha}(\bm{q}, z) dz,
\label{eq:V-prime}
\end{align}

\begin{align}
\tilde{V}_{\delta}^{'' \alpha}(\bm{q}) & =\int \psi_{0 \alpha}(\bm{q}, z)\tilde{V}_{\delta}^{\alpha}(\bm{q})\phi_{\alpha}(\bm{q}, z) dz.
\label{eq:V-2primes}
\end{align}

The approach discussed above, although very rough, nevertheless captures the main physics of the magnetic proximity mechanism acting in our system due to the tunneling of confined spin-polarized charge carriers from the FM $\delta$-layer to the QW. As we see from Eq.~\eqref{eq:E-alpha}, the energy bands $E_{\alpha}^{QW}(\bm{q})$   originating from the nonmagnetic QW states become spin-polarized due to their quantum interference with the spin-polarized states of the FM $\delta$-layer. It can be easily shown that to exponential accuracy, the spin splitting of the states in the QW  $\Delta^{QW}(\bm{q}) = \Delta_{+}^{QW}(\bm{q}) -\Delta_{-}^{QW}(\bm{q})$  contains different contributions decreasing at $L \gg \max\{l_0(\bm {q}), \lambda_{\alpha}(\bm{q})\}$, but the principal  contribution at  $\lambda_{-}(\bm{q}) > l_0(\bm {q}) > \lambda_{+}(\bm{q})$  is  $\Delta^{QW}(\bm{q})\sim \exp[-2L/\lambda_{-}(\bm{q})]$. Note that the lengths $\{l_0(\bm {q}), \lambda_{\alpha}(\bm{q})\}$ decrease at large values of the crystal momentum, $q > \{l_0(0), \lambda_{\alpha}(0)\}^{-1}$ , i.e. corresponding components of $\Delta_{\alpha}^{QW}(\bm{q})$ are exponentially vanishing. That is why, using Eq.~\eqref{eq:E-alpha}, we shall limit ourselves only to  the range of small crystal momenta $q \ll \{l_0(0), \lambda_{\alpha}(0)\}^{-1}$. Qualitative estimates show that for our system, $\lambda_{-}(0) \sim 3-4$ nm, $l_0(0) \sim 1-1.5$ nm, and  $\lambda_{-}(0) \sim 0.5-0.8$ nm, so for the hole densities in QW $n_h^{QW} \sim q_F^2 < 10^{17}$ m$^{-2}$,  our approximation seems to be reasonable (here $q_F$ is the Fermi crystal momentum of holes in the QW).

Thus, we regard the energies given by Eq.~\eqref{eq:E-alpha} (at small wavevectors  $q \ll \{l_0(0), \lambda_{\alpha}(0)\}^{-1}$) as eigenvalues of an effective 2D Hamiltonian $H_{QW}^{eff}$ of charge carriers in the carriers in QW
\begin{align}
H_{QW}^{eff} & = \varepsilon_0(-i\nabla_{\rho}) + U_0 + J_0\bm{e}\cdot \bm{\sigma},
\label{eq:Heff-QW}
\end{align}

\begin{align}
(J_0, U_0) & = \frac{1}{2} [\Delta_{+}^{QW}(0) \mp \Delta_{-}^{QW}(0)],
\label{eq:J-U}
\end{align}
where $\bm{\rho}$ is a 2D  lateral vector in the $(x, y)$ plane, $\bm{e}$ is the unit vector along the arbitrary quantization axis. Hamiltonian~\eqref{eq:Heff-QW} describes 2D carriers in the QW moving in an ``external'' homogeneous field induced by their quantum interference with the spin-polarized carriers in the FM $\delta$-layer. This field has the ``Coulomb'' ($U_0$)  and ``exchange'' ($J_0$) components.

Let us now suppose that due to the lateral long-range
fluctuations of the Mn density within the FM $\delta$-layer on length scales exceeding $\lambda_{-}(0) \sim 3-4$ nm, the components $U_0$ and $J_0$ become slowly varying functions of $\bm{\rho}$ due to the corresponding variation of the potentials $\tilde{V}_{\delta}^{\alpha}$ in Eq.~\eqref{eq:Delta-alpha}.  In other words, we may treat $U_0(\bm{\rho})$ and $J_0(\bm{\rho})$ as slowly varying components of an effective random field projecting the long-range disorder from the FM $\delta$-layer to the QW. We can also include in $U_0(\bm{\rho})$ the ``true'' long-range fluctuating Coulomb potential $\phi(\bm{\rho})$ of charge carriers provided by their inhomogeneous distribution in the $\delta$-layer. The effective random field ($U_0(\bm{\rho}), J_0(\bm{\rho})$) may induce a percolative metal-insulator transition in the QW and the formation of an inhomogeneous structure with the activation-type conductivity. In this situation, the QW can be considered as a set of metallic FM droplets separated by insulating spacers. It is evident from Eq.~\eqref{eq:Heff-QW} that for a hopping between $i$th and $j$th  FM droplets with different orientations ($\bm{e}_i, \bm{e}_j$) of their magnetic moments, a charge carrier in the QW has to overcome an additional energy barrier $J_0[1- (\bm{e}_i \cdot\bm{e}_j)]$ resulting from the effect of the proximity-induced spin polarization.

\section{Models \label{sec:Models}}

In this section, we first of all recall some important methodological aspects of our preceding study~\cite{tripathi11} of the GaAs/In$_{x}$Ga$_{1-x}$As/GaAs QW structure $\delta$-doped by Mn
and discuss existing principal difficulties of an adequate description of magnetism in these multi-component system. Strictly speaking, there is no universal model of magnetism in 2D DMS structures and the choice of a suitable description of FM ordering in such structures is in some aspects more difficult than that in bulk DMSs. For example, let us qualitatively discuss the Zener/RKKY and double exchange pictures of magnetism in the context of the GaAs/In$_{x}$Ga$_{1-x}$As/GaAs QW structure $\delta$-doped by Mn earlier studied in Ref.~\onlinecite{tripathi11}. These structures demonstrate an activation character of conductivity and the key problem is how their FM behavior may be mediated by the holes both in the QW and in the $\delta$-layer. Let us suppose that the effect of ``external'' holes in the QW on the FM ordering in the  $\delta$-layer is less significant than that of ``intrinsic'' holes in the  $\delta$-layer itself. The surface density of manganese ions estimated for the sample studied in the paper (see Sec. IV, sample 4 in Refs.~\onlinecite{tripathi11} is $n_{Mn} \approx 0.4\cdot 10^{18}$ m$^{-2}$), corresponding to the inter-manganese distance of  about 1.6 nm. We consider $n_{Mn}$ as a nominal concentration of local magnetic moments in the $\delta$-layer. Unfortunately, we are unable to directly extract from the conductivity measurements the density of charge carriers (holes) in the $\delta$-layer, $n_h^{\delta}$, due to their low mobility. We can roughly treat $n_{Mn}$ as an upper limit for $n_h^{\delta}$, while the compensation between acceptors and donors in the $\delta$-layer can diminish this estimate by an order of magnitude. Assuming further, that only holes in the QW participate in the conductivity, it is possible to determine a lower limit for  $n_h^{\delta}$ as being of the order of the hole density in the QW, $n_h^{QW}$. Taking $n_h^{QW}\approx 2\cdot 10^{16}$ m$^{-2} \ll n_{Mn}$, from the conductivity data, we obtain the inter-hole distance in the $\delta$-layer $r_h^{\delta}$ falling within the 1.6--7 nm range.

A naive estimate of the hole effective radius in the $\delta$-layer ($a_h^{\delta}$) in the model of a single shallow acceptor gives $a_h^{\delta} \approx a_B \approx 5.3$ nm, where $a_B$ is the Bohr radius of a light hole in GaAs. This value can be treated as an upper limit for $a_h^{\delta}$, while it is obviously overestimated and even exceeds the thickness of the $\delta$-layer. In the more realistic model of a single deep impurity acceptor, we obtain $a_h^{\delta} \approx 0.9$ nm for the heavy hole in GaAs using experimentally known binding energy $E_{\text{Mn}} \approx 90$ meV  for  Mn$^{2+}$ ion in  GaAs. This value can be treated as a lower limit for $a_h^{\delta}$.  So, we get an estimate 0.9 nm  $ < a_h^{\delta} < $ 5.3 nm.

At $r_h^{\delta} <2a_h^{\delta}$, i.e. at a sufficiently high hole density, the Zener/RKKY mechanism is relevant for our system and describes well the FM ordering in the $\delta$-layer within the framework of an effective Heisenberg spin Hamiltonian with an effective coupling via itinerant holes. On the other hand, the magnetism in the polaron model arises by double exchange mechanism which is relevant at a sufficiently low concentration of the holes in the $\delta$-layer, such that $r_h^{\delta} > 2a_h^{\delta}$. In the polaron percolation picture of Kaminski and Das Sarma~\cite{kaminski02}, the holes are localized at the length scale of $a_h^{\delta}$ with strong Hund's rule coupling to Mn atoms.

The above  estimates are {\it a priori} unable to show preference to one or another model of ferromagnetism in our system, due to evident uncertainties in the parameters $r_h^{\delta}$ and $a_h^{\delta}$. Thus, we have to present additional arguments supporting or rejecting the model under consideration. In the following, we analyze the peculiarities of dynamic spin fluctuations in different models and calculate the frequency and time dependence of the resistivity noise power. We show that magnetization dynamics is sensitive to the microscopic nature of FM correlations and thus can be used to distinguish between Zener/RKKY and double-exchange models of FM.

Now, let us come back to the main issue of this paper. In an earlier work, \cite{tripathi11}, we had studied the role of charge disorder and magnetization in the Mn $\delta$-layer on the resistivity of the quantum well. At low charge carrier densities, the random distribution of ionized Mn atoms in the Mn layer creates a fluctuating potential for the holes and leads to accumulation of holes in droplets. Conduction takes place through hopping of holes between the droplets. The building
of magnetic correlations between the charge droplets gives rise to
a dip/shoulder-like feature in the temperature dependence of resistance. The temperature dependence of the resistivity of the droplet phase in these DMS heterostructures was modeled as \begin{align}
\rho(T) & =\rho_{0}e^{\Delta/T+(J_{0}/T)(1-\langle\cos\theta_{ij}
\rangle)},\label{eq:rho}\end{align}
 where $\theta_{ij}$ is the angle between the orientations of the
magnetizations $\mathbf{S}_{i}$ and $\mathbf{S}_{j}$ associated
with the neighboring $i^{th}$ and $j^{th}$ droplets, respectively,
and $\langle\cos\theta_{ij}\rangle\propto\langle\mathbf{S}_{i}
\cdot\mathbf{S}_{j}\rangle.$
The temperature dependence of the static magnetic correlation for
a 2D ferromagnet above the transition temperature will in general
take the form \begin{align*}
\langle\mathbf{S}_{i}\cdot\mathbf{S}_{j}\rangle & \sim e^{-R_{ij}/\xi_{M}(T)},\end{align*}
 where $\xi_{M}$ is the magnetic correlation length. The static magnetic
susceptibility $\chi_{0}$ is related to $\xi_{M}$ through $\chi_{0}(T)\sim T\xi_{M}^{2}.$
The two mechanisms that are generally considered for ferromagnetism
in DMS systems are double exchange and $p$-$d$ Zener/RKKY. The low-energy excitations of both these models are known to have same the dispersion relation $\omega_{\mathbf{q}}\propto q^{2}$ and as far as static properties are concerned, both can be represented by an effective Heisenberg model of spins at low energies. Therefore, the two models predict the same behavior of the resistivity and cannot distinguish between the two mechanisms of ferromagnetism in the Mn layer. For a Heisenberg model of spin interactions, 
\begin{align}
H_{\text{Heisen}} & =-\sum_{\langle ij\rangle}J_{ij}\mathbf{S}_{i}\cdot\mathbf{S}_{j},
\label{eq:heisenberg}\end{align}
the magnetic correlation length is given by
\begin{align}
\xi_{M}(T) & \sim a\sqrt{JS/T}e^{\pi T_{C}/2T}. \label{eq:xi-heisenberg}\end{align}
 The real system is likely to deviate from an isotropic Heisenberg
ferromagnet due to spin-orbit interactions as well as the 2D distribution of the Mn doping. For instance, for a uniaxial ferromagnet, we have
\begin{align}
H_{\text{Heisen}} & =-J\sum_{\langle ij\rangle}\mathbf{S}_{i}\cdot\mathbf{S}_{j}-K\sum_{i}
(S_{iz})^{2}.\label{eq:model-H}\end{align}
 For this uniaxial magnet $S_{z}$ is a conserved order parameter.
For small but finite $K>0,$ the model shows Ising-type behavior at
sufficiently low temperatures, undergoing a transition to a ferromagnetically
ordered state \cite{bander88} at a temperature $T_{0}\ll T_{C}:$
\begin{align}
T_{0} & \sim\frac{T_{C}}{\ln(\pi^{2}J/K)}.\label{eq:ising-trans}\end{align}
 For temperatures $T_{C}\gg T\gg T_{0},$ the correlation length is
approximately given by Eq.~\eqref{eq:xi-heisenberg}. As the temperature approaches $T_{0},$ Eq. (\ref{eq:xi-heisenberg}) for the magnetic correlation length should perhaps be replaced with a power law,
\begin{align}
\xi_{M} & \sim a/(T/T_{0}-1)^{\gamma},
\label{eq:xi-ising}\end{align}
where $\gamma=1.25$ for the limiting case of the Ising model. For
a general anisotropic Heisenberg model, the order parameter is not
conserved. In Ref. \onlinecite{tripathi11}, we considered an isotropic Heisenberg model and obtained quantitative agreement with the resistivity of the insulating samples over a broad temperature range.

As is clear from the above discussion, resistivity measurements in
these DMS structures are unable to distinguish between RKKY and double-exchange mechanisms since the static correlations in both cases can be described by an equivalent effective Heisenberg model. It is then natural to examine dynamic properties such as the (frequency-dependent) resistivity noise. Resistivity noise can of course also arise from charge fluctuations in the puddles in the quantum well. However, if noise from magnetic fluctuations dominates that from charge fluctuations, resistivity noise can be a useful tool for probing the microscopic origin of ferromagnetic interactions. This happens to be true in our case.

We develop now a theory for the effect of magnetic and charge fluctuations on the resistivity noise. We will also examine the effect of various kinds of disorder on the noise.

\subsection{Resistivity noise from magnetic fluctuations}

We consider the magnetic fluctuations first. We will study the implication of increasing magnetic anisotropy on the frequency dependence of the resistivity noise.

For homogeneously disordered magnets, it is known that the resistivity noise $S_{\rho}(\omega)=\langle|\delta\rho(\omega)|^{2}\rangle$ and
the magnetization noise $S_{M}(\omega)=\langle|\delta M(\omega)|^{2}\rangle$
are related through \cite{hardner93} $S_{\rho}(\omega)=S_{M}(\omega)(d\rho/dM)^{2}.$
A similar relation can be obtained for our phase-segregated model.
We assume that the magnetic moments in a droplet are aligned with
the polarization in the Mn layer directly above the droplet and disregard fluctuations of the magnitude of the droplet magnetic moment. Our model for resistivity is a simple nearest-neighbor hopping type
and the hopping is a function of the relative orientation of the magnetizations
at the sites of the two puddles in question: $\rho\equiv\rho_{ij}(\mathbf{M}_{i}\cdot\mathbf{M}_{j})=\rho_{ij}(\sum_{\alpha}M_{i}^{\alpha}M_{j}^{\alpha}).$ Fluctuations of the orientation of the magnetizations of the droplets cause resistivity fluctuations. For our model of resistivity, $\delta\rho/\delta M_{i}^{\alpha}=-(J_0/T)\rho M_{j}^{\alpha}.$ We thus obtain $\delta\rho=\sum_{\alpha}(\frac{\delta\rho}{\delta M_{i}^{\alpha}}\delta M_{i}^{\alpha}+\frac{\delta\rho}{\delta M_{j}^{\alpha}}\delta M_{j}^{\alpha})=-(J_0/T)\sum_{\alpha}\rho(M_{i}^{\alpha}\delta M_{j}^{\alpha}+M_{j}^{\alpha}\delta M_{i}^{\alpha}).$
Here $\rho$ and $M_{k}^{\alpha}$ refer to the static values, and
the time dependence is expressed in the fluctuations $\delta M_{k}^{\alpha}.$ The Greek labels refer to the orientation of the magnetization and the Latin indices label the charge puddles in the quantum well. $i$ and $j$ in the above expressions refer to nearest-neighbor puddles in the quantum well and are not summed over. 
The resistivity noise takes the form
\begin{align}
S_{\rho}(\omega) & =\sum_{ij,\alpha\beta}\left(\frac{\delta\rho}{\delta M_{i}^{\alpha}}\right)\left(\frac{\delta\rho}{\delta M_{j}^{\beta}}\right)C_{\alpha\beta}(R_{ij},\omega),
\label{eq:S-rho}\end{align}
where $C_{\alpha\beta}(R_{ij},\omega)=\int dt\, e^{i\omega t}\langle\delta M_{i}^{\alpha}(t)\delta M_{j}^{\beta}(0)\rangle$
is the correlation function of the magnetization, $\alpha,\beta$
refer to magnetization directions and $i,j$ refer to neighboring
droplets. From Eqs.~(\eqref{eq:rho}) and (\eqref{eq:S-rho}) it thus follows that
\begin{align}
S_{\rho}(\omega)/\rho^{2} \sim \quad \quad \quad \quad \quad    \nonumber \\ (J_{0}/T)^{2}\sum_{\alpha}\left[C_{\alpha\alpha}
(0,\omega)+\langle\cos\theta_{ij}\rangle C_{\alpha\alpha}(R_{ij},\omega)\right]\nonumber \\
\approx(J_{0}/T)^{2}\sum_{\alpha}\left[C_{\alpha\alpha}
(0,\omega)+e^{-R_{ij}/\xi_{M}(T)}C_{\alpha\alpha}(R_{ij},\omega)
\right].\label{eq:r-noise}\end{align}
The spin correlation function is related to the dynamic susceptibility through the fluctuation-dissipation relationship,
\begin{align*}
C_{\alpha\beta}(\mathbf{q},\omega) & =\frac{2T}{\omega}\text{Im }\chi_{\alpha\beta}(\mathbf{q},\omega).\end{align*}

\subsection{Resistivity noise for different models of magnetism \label{sub:Resistivity-noise-for}}

The magnetization dynamics is crucially dependent on whether the order parameter is a conserved quantity. If the order parameter is not conserved, then the spin relaxation has a finite lifetime even as $q \rightarrow 0.$ This is the case for an Ising or anisotropic Heisenberg ferromagnet.
If the order parameter is conserved, then the spin relaxation lifetime diverges as $q\rightarrow0.$ This would be the case, for example, for a uniaxial or isotropic Heisenberg ferromagnet. We follow the hydrodynamic approach of Hohenberg and Halperin \cite{hohenberg77} for low energy dynamics of all these cases.

Consider first the case where the order parameter is not a conserved
quantity. This could be the case, for example, for anisotropic magnets. Such a case corresponds to Model A of Ref.~\onlinecite{hohenberg77}, defined by the Markoffian equations of motion
\begin{align}
\frac{\partial\psi_{\alpha}(\mathbf{r},t)}{\partial t} & =-\Gamma_{0}\frac{\delta F_{0}}{\delta\psi_{\alpha}(\mathbf{r},t)}+\theta_{\alpha}
(\mathbf{r},t)+\Gamma_{0}h_{\alpha}(\mathbf{r},t),\nonumber \\
F_{0} & =\int d\mathbf{r}\left[\frac{1}{2}\xi_{M}^{-2}\psi^{2}+\frac{1}{2}
|\nabla\psi|^{2}+u_{0}\psi^{4}\right],\nonumber \\
\psi^{2} & =\sum_{\alpha=1}^{n}\psi_{\alpha}^{2};\quad\psi^{4}=
(\psi^{2})^{2}.\label{eq:model-A}\end{align}
 $\theta_{\alpha}(\mathbf{r},t)$ is a Gaussian white noise source,\begin{align}
\langle\theta_{\alpha}(\mathbf{r},t)\rangle & =0,\nonumber \\
\langle\theta_{\alpha}(\mathbf{r},t)\theta_{\alpha'}
(\mathbf{r}',t')\rangle & =2\Gamma_{0}\delta(\mathbf{r}-\mathbf{r}')\delta(t-t')
\delta_{\alpha\alpha'},\label{eq:theta-noise}\end{align}
 and $h_{\alpha}(\mathbf{r},t)$ are arbitrary external fields. The
dynamic susceptibility for this model has the form\begin{align}
\chi_{\alpha\beta}^{(A)}(\mathbf{q},\omega) & =\frac{\chi_{0}(T)\delta_{\alpha\beta}}{1+(q\xi_{M})^{2}
-i\omega\xi_{M}^{2}/\Gamma_{0}}.
\label{eq:chi-A}\end{align}
 The constant $\Gamma_{0}$ will in general depend on $\xi_{M}$ and
$\mathbf{q},$ but the relaxation rate $\Gamma_{0}(\mathbf{q})/\xi_{M}^{2}=\text{const.}$
as $\mathbf{q}\rightarrow0$ \cite{hohenberg77}. At finite temperatures, the autocorrelation function will exhibit an exponential relaxation with
time,\begin{align*}
C_{\alpha\beta}(R_{ij},t) & \sim\delta_{\alpha\beta}Te^{-\Gamma_{0}t/\xi_{M}^{2}-R_{ij}/\xi_{M}}.
\end{align*}
The resistivity noise from magnetic fluctuations is of the random
telegraph type:
\begin{align}
\langle|\delta\rho_{\omega}|^{2}\rangle/\rho^{2} & \sim J_{0}^{2}\xi_{M}^{2}/T\Gamma_{0}\frac{1+e^{-2R_{ij}/\xi_{M}}}
{(\omega\xi_{M}^{2}/\Gamma_{0})^{2}+1}.\label{eq:r-noise2}\end{align}

Consider now the case where the order parameter is conserved (Model
B in the parlance of Ref.~\onlinecite{hohenberg77}). This would
be the case, for example, for a uniaxial or isotropic ferromagnet.
The dynamical susceptibility has a form similar to Eq.~\eqref{eq:chi-A}, except that now spin relaxation lifetime diverges as $q\rightarrow0$ so that $\chi(0,\omega)=0.$ The RKKY and double exchange models as well as the pure Heisenberg model model fall in this category. For a pure Heisenberg model, the damping mechanism is magnon-magnon scattering and one has $\Gamma_{0}(\mathbf{q})/\xi_{M}^{2}\propto q^{2}$. Thus, in this case \begin{align}
\chi_{\alpha\beta}^{(B)}(\mathbf{q},\omega) & =\frac{\chi_{0}(T)\delta_{\alpha\beta}}
{1+(q\xi_{M})^{2}-i\omega/Dq^{2}}.
\label{eq:chi-B}\end{align}
The autocorrelation function $C_{\alpha\beta}(R_{ij},\omega)$ is
given by
\begin{align}
C_{\alpha\beta}(R_{ij},\omega) & =\delta_{\alpha\beta}\frac{T\chi_{0}}{\pi}\int_{0}^{\infty}dq\, q\frac{Dq^{2}J_{0}(qR_{ij})}{(Dq^{2}(1+(q\xi_{M})^{2})^{2}
+\omega^{2}}\nonumber \\
 & =\delta_{\alpha\beta}\frac{T\chi_{0}}{\pi D}\int_{0}^{\infty}dy\frac{y^{3}J_{0}(yR_{ij}/\xi_{M})}
{(y^{2}(1+y^{2}))^{2}+(\omega\xi_{M}^{2}/D)^{2}}.
\label{eq:C-B-w}\end{align}
 For separations $R_{ij}$ such that $(\omega\xi_{M}^{2}/D)^{1/4}R_{ij}/\xi_{M}\ll1,$
Eq.~\eqref{eq:C-DE-w} has the following limiting behavior: \begin{align}
 C_{\alpha\beta}(0,\omega) \approx \quad  \quad \nonumber
\\
\delta_{\alpha\beta}\frac{T\chi_{0}}{2\pi D}\left\{
\begin{array}{c} \ln(D/\omega\xi_{M}^{2})-1+\frac{3\pi}{2}
(\omega\xi_{M}^{2}/D),\quad\omega\xi_{M}^{2}/D\ll1\\
\frac{\pi}{4}(D/\omega\xi_{M}^{2})-\frac{\pi}
{4\sqrt{2}}(D/\omega\xi_{M}^{2})^{2},
\quad\omega\xi_{M}^{2}/D\gg1\end{array}.
\right.\label{eq:C-B-w-lim}\end{align}
Equation~(\eqref{eq:C-B-w-lim}) should be contrasted with the Lorentzian
behavior of the autocorrelation function for random telegraph noise.
It is also useful to study the time dependence of the autocorrelation function,
\begin{align}
C_{\alpha\beta}(R_{ij},t) & =\delta_{\alpha\beta}\frac{\chi_{0}}{4\pi}\int dq\, q\frac{e^{-Dq^{2}(1+(q\xi_{M})^{2})t}J_{0}(qR_{ij})}
{1+(q\xi_{M})^{2}}.\label{eq:C-B-t}
\end{align}
At small separations $R_{ij}$ such that $(R_{ij}/\xi_{M})(\xi_{M}^{2}/Dt)^{1/4}\ll1,$ we may ignore the Bessel function in Eq.~\eqref{eq:C-B-t} and obtain
\begin{align}
C_{\alpha\beta}(0,t) & \sim\delta_{\alpha\beta}\frac{\chi_{0}}{\xi_{M}^{2}}\left\{ \begin{array}{c}
\ln(\xi_{M}^{2}/Dt),\quad Dt/\xi_{M}^{2}\ll1\\
(\xi_{M}^{2}/Dt)-\frac{1}{3}(\xi_{M}^{2}/Dt)^{2},\quad Dt/\xi_{M}^{2}\gg1\end{array}.\right.\label{eq:C-B-t-lim}\end{align}

Next, we consider the RKKY mechanism in a clean metal. In this case,
the magnons can decay into particle-hole excitations for which one
can show~\cite{mahan} $\Gamma_{0}(\mathbf{q})\sim\gamma q.$ However,
in the presence of impurities, one must take into account diffusion
corrections~\cite{finkelstein84} which results in
\begin{align}
\chi_{\alpha\beta}^{\text{RKKY}}(\mathbf{q},\omega) & \approx\frac{\chi_{s}(T)\delta_{\alpha\beta}}{1+(q\xi_{M})^{2}-i
\omega/D_{s}q^{2}},\label{eq:chi-rkky}\end{align}
 where $\chi_{s}$ and $D_{s}$ are the uniform spin susceptibility
and spin diffusion constant respectively for the disordered system.
Note that this gives us the same results for the autocorrelation function as the pure Heisenberg model with conserved spin dynamics. The mechanism of damping is however different - here it is magnon decay into incoherent particle-hole excitations.

Finally let us consider the double exchange model for which the spin
wave life time is given by $\Gamma_{0}(\mathbf{q})/\xi_{M}^{2}\propto q^{5}$ in two dimensions \cite{golosov2000,shannon02}. For this model, we have
\begin{align}
\chi_{\alpha\beta}^{\text{DE}}(\mathbf{q},\omega) & =\frac{\delta_{\alpha\beta}\chi_{0}^{\text{DE}}(T)}
{1+(q\xi_{M})^{2}-i\omega/Aq^{5}}.\label{eq:DE-susc}\end{align}
The autocorrelation function $C_{\alpha\beta}(R_{ij},\omega)$ is
given by
\begin{align}
C_{\alpha\beta}(R_{ij},\omega) & =\delta_{\alpha\beta}\frac{T\chi_{0}}{\pi}\int_{0}^{\infty}dq\, q\frac{Aq^{5}J_{0}(qR_{ij})}{(Aq^{5}(1+(q\xi_{M})^{2})^{2}
+\omega^{2}}\nonumber \\
& =\delta_{\alpha\beta}\frac{T\xi_{M}^{3}\chi_{0}}{\pi A}\int_{0}^{\infty}dy\frac{y^{6}J_{0}(yR_{ij}/\xi_{M})}
{(y^{5}(1+y^{2}))^{2}+(\omega\xi_{M}^{5}/A)^{2}}.
\label{eq:C-DE-w}\end{align}
For small enough $R_{ij},$ the autocorrelation function for the
double exchange ferromagnet has the following limiting behavior
\begin{align}
C_{\alpha\beta}(0,\omega) \approx \quad \quad \quad \quad \nonumber
\\
\delta_{\alpha\beta}\frac{T\xi_{M}^{3}\chi_{0}}{\pi A}\left\{ \begin{array}{c}
\frac{\pi}{8\cos(\pi/5)}\frac{1}{(\omega\xi_{M}^{5}/A)^{3/5}},
\quad\omega\xi_{M}^{5}/A\ll1\\
\frac{\pi}{14}\frac{1}{(\omega\xi_{M}^{5}/A)},
\quad\omega\xi_{M}^{5}/A\gg1\end{array}.\right.
\label{eq:C-DE-w-lim}\end{align}
Thus, at long times, the autocorrelation function (and consequently
the resistivity noise) for the 2D double exchange ferromagnet decays
as $C_{\alpha\beta}(0,t)\sim1/t^{2/5}.$ This is to be contrasted
with the $1/t$ decay for the Heisenberg and disordered RKKY ferromagnets. As we shall see in the following, the experimental data on resistivity noise is consistent with the RKKY model and not the double exchange model.

\subsection{Disorder effects}

In Sec. \ref{sub:Resistivity-noise-for}, we considered the effect
of disorder on the dynamic susceptibility of the holes in the Mn layer. We found that unlike the case of a clean metal where $\Gamma_{0}(\mathbf{q})\sim\gamma q,$ weak localization effects give us $\Gamma_{0}(\mathbf{q})\sim D_{s}q^{2}$ instead. Thus, the magnetization dynamics of disordered RKKY magnets incorporating weak localization effects and a purely Heisenberg magnet are the same. In this section, we discuss the effects of disorder arising from randomness in the positions of the Mn spins. We model such disorder
in the form of site dilution and randomly varying exchange interaction, respectively, starting from a uniform Heisenberg model.

First, we consider a nearest-neighbor Heisenberg model with a fraction $c=1-p$ of
missing sites, which mimics random concentration of Mn atoms.  $c=1-p$ does not refer to the monolayer doping density of Mn atoms; rather, it refers to missing sites on an effective sublattice made of the Mn atoms with a lattice constant of the order of the typical inter-Mn separation. The exchange energy scale in the model is set by the value at typical Mn-Mn separations. A missing site can be related to regions with no Mn atoms on the sublattice. For $p>p_{c}$ (the critical threshold for percolation), the ground state is ferromagnetic and the low-energy spin dynamics is the same as that for the pure Heisenberg model, albeit with a reduced zero-temperature spin-wave stiffness.\cite{harris77}. The temperature dependence of the magnetic correlation length for $T\ll T_{C}$ is thus $\xi_{M}\sim\exp[\pi T_{C}(p)/2T].$\cite{aronovitz90} In particular, for the 2D isotropic model on a square lattice, $T_{C}(p)\sim(p-p_{c})^{1.296}$ for $p\rightarrow p_{c},$ \cite{aronovitz90} and  $T_{C}(p)\propto1.33p^{2}-0.33$
for $p\sim1.$\cite{harris77} Thus disorder due to site dilution
will not affect the frequency and temperature dependences of noise
for $T\ll T_{C}$ as long as one remains above the site percolation threshold on this sublattice. In actual systems, the long range of exchange interactions means that the percolation transition may not occur and one remains in the Heisenberg universality class even for random concentration changes.

Next, we consider a nearest-neighbor Heisenberg model with random exchange interactions chosen with zero mean. More realistically, the exchange interactions between pairs of spins will be a rapidly oscillating function of their separation. However, for the spin-glass state we are describing, the long-range nature of the interaction is not believed to play an important role and we therefore discuss the nearest-neighbor Heisenberg model with couplings of random sign as opposed to the actual system with random site positions. The low-energy excitations of this model are gapless linearly-dispersing magnons\cite{halperin77} (as in a Heisenberg antiferromagnet), and the lower critical dimension is two. The transverse susceptibility at low temperatures is expected to behave like $\chi(T)\sim e^{\pi T_{C}/T}$ as is the case for the Heisenberg ferromagnet/antiferromagnet models. The dynamics of the spin-glass depends on whether one has dissipative equations of motion or the total spin is conserved.\cite{hertz79} For dissipative dynamics,
Model A discussed above is evidently appropriate while for conserved
dynamics, the spin dynamics is diffusive (like in Model B).~\cite{halperin77,hertz79} To summarize, disorder in the form of site dilution (while remaining above the percolation threshold) as well as in the form of a rapidly oscillating function of the inter-Mn separation does not qualitatively change the results for Model A and Model B dynamics discussed above for pure systems.

Finally, let us consider the case where the disordered magnetic system is in a Griffiths phase. At a finite temperature, the system is organized into clusters that are weakly coupled to their neighbors but within the clusters, the spins are ferromagnetically ordered
(If the exchange coupling $J$ to a neighboring spin is less than the temperature $T,$ we may regard the coupling to be weak).
The long-time behavior of the autocorrelation function is dominated by the slow relaxation of atypically large, compact clusters. The relaxation of the cluster spin is through spin diffusion. The probability that a site $i$ belongs to a cluster of size $L$ is of the order of $e^{-c(L/\xi)^{2}},$ where $\xi$ is the magnetic correlation length, and $c$ is a constant. The autocorrelation function, $C(t)\sim\sum_{L}e^{-c(L/\xi)^{2}-{\cal D'}t/L^{2}}$ is dominated by contributions from clusters of size $L_{*}^{2}=\xi\sqrt{{\cal D}'t/c},$ whence~\cite{bray87} \begin{align*}
C(t) & \sim e^{-2\sqrt{{\cal D}'ct}/\xi}.
\end{align*}
The resistivity noise at low frequencies has the following behavior
\begin{align}
\langle|\delta\rho_{\omega}|^{2}\rangle/\rho^{2} & \sim(J_{0}^{2}\xi^{2}/{\cal D}'c)(1+e^{-2R_{ij}/\xi}).\label{eq:r-griffiths}\end{align}

\subsection{{Resistivity noise from charge fluctuations}\label{sec:charge-fl}}

Resistivity noise can arise from charge fluctuations due to inter-droplet hopping. The effect of inter-droplet hopping on the resistivity noise has been studied in the literature in the context of electronic phase separation in manganites~\cite{rakhmanov01}, and we make a similar analysis for our droplet phase. We begin with the ground state consisting of a large number $N$ of neutral (uncharged) droplets. At finite temperatures, let $N_{0}$ be the number of neutral droplets and $N_{\pm}$ be the number of droplets with one extra (less) charge. Evidently $N_{+}=N_{-}$ and $N=N_{0}+2N_{+}.$ Let $E_{C}$ denote the charging energy of the droplets, and $\omega_{0}$ be the rate of escape of a hole from a charge-neutral droplet. In equilibrium, $N_{+}=Ne^{-E_{C}/T}.$
There are four elementary charge transfer processes - (1) from one
neutral droplet to another, (2) from a droplet with one excess charge to another with one deficient charge, (3) from a droplet with one
excess charge to a neutral droplet, and (4) from a neutral droplet
to another with one deficient charge. The latter two processes do
not involve a free energy cost. At low temperatures, we may ignore
processes involving transfers of larger charges. The rates for processes (1) and (2) are, respectively,
\begin{align}
\frac{1}{\tau_{1,2}(\mathbf{r}_{ij})} & =\omega_{0}e^{-r_{ij}/\xi_{\text{loc}}\mp E_{C}/T},\label{eq:tau12}\end{align}
where $r_{ij}$ is the distance between the two droplets. For processes (3) and (4) one has
\begin{align}
\frac{1}{\tau_{3,4}(\mathbf{r}_{ij})} & =\omega_{0}e^{-r_{ij}/\xi_{\text{loc}}}.
\label{eq:tau34}\end{align}
In the presence of magnetic correlations, we should replace $\omega_{0}$ by $\omega_{M}=\omega_{0}e^{-(J_{0}/T)(1-\cos\theta_{ij})}.$ Similar
considerations also apply for the fluctuations of the number of charged and neutral droplets; for example, for process (2), $\delta\dot{N}_{+}(t,\mathbf{r}_{i})=-\sum_{j}\delta N_{+}(t,\mathbf{r}_{i})/\tau_{2}(\mathbf{r}_{ij}),$
where the summation is over all droplets with one deficient charge.
The average fluctuation of $N_{+}$ is $\langle\delta N_{+}^{2}\rangle_{T}=N_{+}/2.$

Resistivity fluctuations are related to fluctuations of $N_{+}$ through the following relationship~\cite{rakhmanov01}:
\begin{align*}
\frac{\delta\rho}{\rho} & =-\frac{\delta N_{+}}{N_{+}}(1-2e^{-E_{C}/T}).
\end{align*}
This leads us to
\begin{align}
\frac{\langle|\delta\rho_{\omega}|^{2}\rangle}{\rho^{2}} & =\langle\delta N_{+}^{2}\rangle_{T}\frac{1-2e^{-E_{C}/T}}{N_{+}^{2}}\sum_{j}
\frac{2\tau_{2}(\mathbf{r}_{ij})}{1+(\omega\tau_{2}
(\mathbf{r}_{ij}))^{2}}\nonumber \\
 & =(1-2e^{-E_{C}/T})\sum_{j}\frac{2\tau_{2}(\mathbf{r}_{ij})}
{1+(\omega\tau_{2}(\mathbf{r}_{ij}))^{2}},
\label{eq:chargfluc1}\end{align}
where the sum $ij$ is over all pairs of droplets corresponding to
process (2). The dominant contribution is from relaxation to nearest
neighbor droplets. This is especially true for the insulating samples where the inter-droplet distance $R>\xi_{\text{loc}}.$ Retaining
only the nearest neighbor contributions, the normalized resistivity
noise due to droplet charge fluctuations takes the form \begin{align}
\frac{\langle|\delta\rho_{\omega}|^{2}\rangle}{\rho^{2}} & \approx z(1-2e^{-E_{C}/T})\frac{2\tau_{2}(R)}{1+(\omega\tau_{2}(R))^{2}},
\label{eq:chargeflucnoise}\end{align}
where $z$ is the droplet coordination number and
 \begin{align}
\frac{1}{\tau_{2}(R)} & \approx\omega_{0}e^{-R/\xi_{\text{loc}}+E_{C}/T-(J_{0}/T)
[1-\exp(-R/\xi_{M})]}.\label{eq:tau2M}\end{align}
The temperature dependence of $\tau_{2}(R)$ is different from that
of the resistivity [see Eq.~\eqref{eq:rho}] but nevertheless shows
an anomalous behavior around a temperature where $\xi_{M}(T)\sim R.$
In the low temperature limit where $R/\xi_{M}\ll1,$ $\tau_{2}(R)\sim e^{-E_{C}/T}$. In the high temperature regime, we have $\tau_{2}(R)\sim e^{-(E_{C}-J)/T}.$

\section{Comparison with experiment \label{sec:Comparison-with-experiment}}

The measurements of the resistivity noise were performed with the sample, the structure of which is presented in Fig.~\ref{fig:structure}. Parameters of this sample are as follows: Mn content is 0.35 ML, and it corresponds to $2\cdot10^{14}$ cm$^{-2}$ surface concentration; the thickness of the spacer separating Mn $\delta$-layer and  the QW is 3 nm; indium concentration in the Ga$_{1-x}$In$_x$As QW $x$=0.17 and so its depth is about 70 meV; the hole density $p$ and their mobility $\mu_p$ in QW are $1.4\cdot10^{12}$ cm$^{-2}$ (77 K) or $0.5\cdot10^{12}$ cm$^{-2}$ (5 K) and 2370 cm$^2$/V$\cdot$s (77 K) or 3400 cm$^2$/V$\cdot$s (5 K), respectively. As it was shown Ref.~\onlinecite{tripathi11}, this sample is close to the metal-insulator transition of the percolation type having high enough value of resistance $R(5 K) = 19.7$ k$\Omega$ and $R(5 K)/R(70 K) \approx 1.5$. The temperature dependence of its resistivity can be fitted to the Arrhenius law with an activation energy $\approx
 20$ K. Ferromagnetic ordering is established for this sample through observations of the anomalous Hall effect and resistivity anomaly (peak at the temperature dependence of the resistance). The direct magnetic measurements for samples from the same set show magnetic hysteresis even for samples with smaller Mn content.~\cite{aronzon07} The temperature corresponding to the onset of strong ferromagnetic correlations for this sample was found to be about 30 K as measured by the resistivity anomaly. As it was shown previously~\cite{tripathi11}, the Mn $\delta$-layer consists of FM droplets affecting the magnetic state in the QW giving rise to a set of metallic FM droplets, which is responsible for metal-insulator transition. So, there are two characteristic temperatures: first that corresponds to the FM ordering inside droplets and the second related to the formation of the long-range FM state in the sample, which can be treated as a kind of the ``FM percolation transition''. The characteristic temperature found from the resistive anomaly corresponds to the second case. At this temperature magnetic correlations became significant.

\subsection{{Frequency and time dependence}\label{sub:freq-time-dep}}

\begin{figure}
\includegraphics[clip,width=0.95\columnwidth]{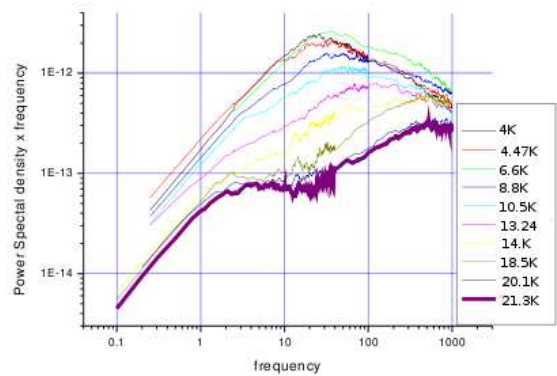} \caption{\label{fig:freq-dep} (color online) Plots of the measured frequency dependence of $fS_{\rho}(f)=f\langle|\delta\rho_{f}|^{2}\rangle$ for various values of the temperature. }
\end{figure}

The data (see Fig.~\ref{fig:freq-dep}) on the frequency dependence
of noise show $fS_{\rho}(f)=f\langle|\delta\rho_{f}|^{2}\rangle$
($\omega=2\pi f$) as a function of frequency for a number of temperatures (mostly below the resistivity peak). The noise is non-monotonous at all these temperatures, and especially so at lower temperatures. The curves at higher temperatures suggest more than one relaxation time while the low temperature data indicates a single relaxation time.

\begin{figure}
\includegraphics[width=0.95\columnwidth]{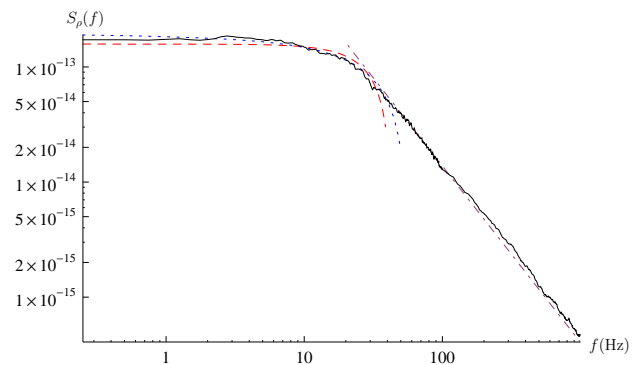}
\caption{\label{fig:freq-dep-4K} (color online) Frequency dependence of noise at $T=4K$ (solid curve) together with fits to the low and high frequency regimes. At the low-frequency end, the dashed curve and the dotted curve are fits to $S_{\rho}\sim A-Bf^{2}$ and $S_{\rho}\sim A-B\ln f-Cf,$ respectively. At the high-frequency end, the fit is to $S_{\rho}\sim Af^{-1.53}.$}
\end{figure}

The data in Fig.~\ref{fig:freq-dep} correspond to temperatures well
below the resistivity anomaly. This is the regime where ferromagnetic correlations are significant. Note that the resistivity noise amplitude decreases with increase of temperature for the high frequency as well as the low-frequency limits. Thus, we try power-law fits in the low- and high-frequency range away from the peak in the resistivity noise (see Fig.~\ref{fig:freq-dep-4K}). This behavior is consistent with both the magnetic and charge fluctuation models. We will discuss the temperature dependence of resistivity noise later in Sec.~\ref{sub:Temperature-dependence}.

We consider power-law fits at the low- and high-frequency ends. At
high frequencies, we find (see Fig.~\ref{fig:freq-dep-4K}) for example for the $T=4K$ data, $S_{\rho}(f)\sim f^{-1.5},$ which should be contrasted with a $f^{-2}$ decrease expected for a Lorentzian. At
low frequencies, one can find a fit to $S_{\rho}(f)\sim A-Bf^{2}$
or $S_{\rho}(f)\sim A-B\ln f-Cf.$ The former would be consistent
with a Lorentzian (no conserved order parameter) while the logarithmic behavior in the latter can arise from Model B dynamics (conserved order parameter).

\begin{figure}
\includegraphics[width=0.95\columnwidth]{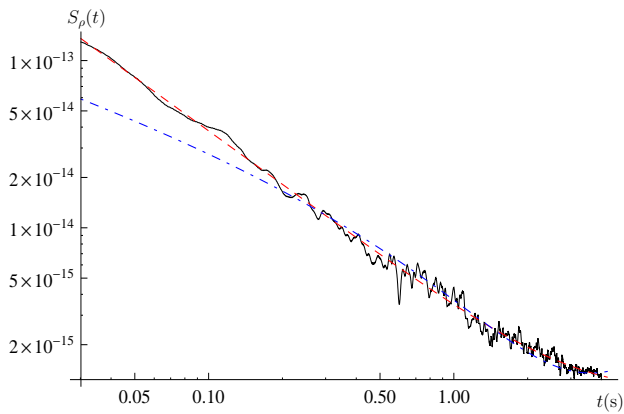}
\caption{\label{fig:griffiths4.4K} (color online) Plot showing the long time dependence of the resistivity autocorrelation function $S_{\rho}(t)$ extracted from the resistivity noise data at $T=4.0{\rm K}$ (solid curve) together with fits to a power-law time dependence with logarithmic corrections. The dashed curve is a fit to $S_{\rho}(t)=A/t^{1.05}+B\ln(t/t_{0}),$ while the dot-dashed curve is a fit to $S_{\rho}(t)=A/t^{2/5}+B\ln(t/t_{0}).$ In two dimensions, $S_{\rho}(t)\sim t^{-1}$ behavior is expected for a disordered RKKY ferromagnet (see Eq.~\eqref{eq:chi-rkky}) and $S_{\rho}(t)\sim t^{-2/5}$ for double exchange ferromagnets (see Eq.~\eqref{eq:DE-susc}). The logarithmic time dependence indicates
$1/f$ noise contributions. The fit to the RKKY model is better than
to the double exchange.}
\end{figure}

By analyzing the long-time behavior of $S_{\rho}(t),$ we find that
the agreement is better with Model B dynamics. Fig.~\ref{fig:griffiths4.4K} shows the time dependence of the resistivity autocorrelation function numerically obtained by a discrete Fourier transform of the frequency data. The long-time behavior shows a good fit to $S_{\rho}(t)\sim A/t^{1.05}+B\ln(t/t_{0}).$ For disordered RKKY ferromagnets, one expects $S_{\rho}(t)\sim1/t,$ while for the double exchange case, $S_{\rho}(t)\sim1/t^{2/5}.$ We found that fits to exponential relaxation, $S_{\rho}\sim e^{-t/\tau}$ or a Griffiths relaxation, $S_{\rho}\sim e^{-\sqrt{t/\tau}}$ are not as good as the power-law fits. The $1/t$ power-law relaxation supports the case of the RKKY ferromagnet.

\begin{figure}
\includegraphics[width=0.95\columnwidth]{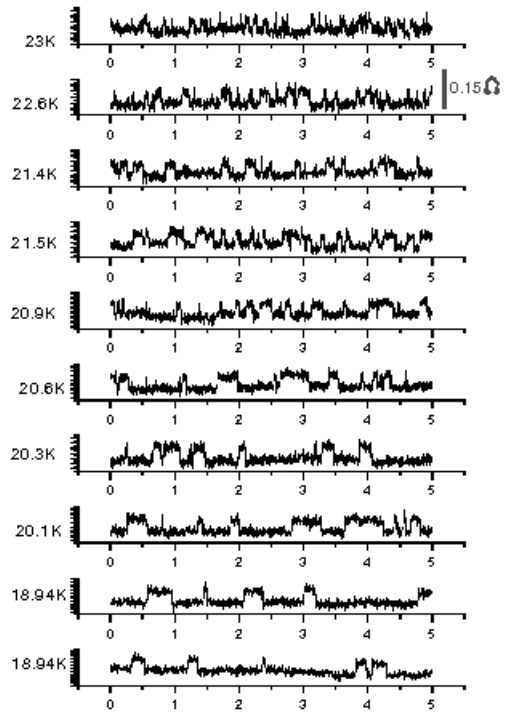}
\caption{\label{fig:telegraph-noise}Time traces of resistivity fluctuations showing distinct telegraph-like switching at low temperatures. At temperatures higher than around 20K, the telegraph-like switching is not so prominent.}
\end{figure}

In Fig.~\ref{fig:telegraph-noise}, the time traces of resistivity fluctuations are shown, while the main stable contribution to the resistivity is subtracted. Note that the time traces of the resistivity show distinct telegraph-like switching at low temperature. This switching is not so prominent once the temperature exceeds about 200 K: this is near to but below the temperature, at which one observes the anomalous peak in the resistivity and noise. The advent of telegraph-like switching can be understood if the magnetic model has uniaxial symmetry instead of complete rotational symmetry. This is possible given the planar geometry of the structure. Nevertheless, at the moment we are unable to say whether at still longer times, the autocorrelation relaxation will remain a power law or become exponential. The logarithmic term indicates $1/f$ noise, possibly from other mechanisms.

\subsection{Temperature dependence \label{sub:Temperature-dependence}}

\begin{figure}
\includegraphics[clip,width=0.95\columnwidth]{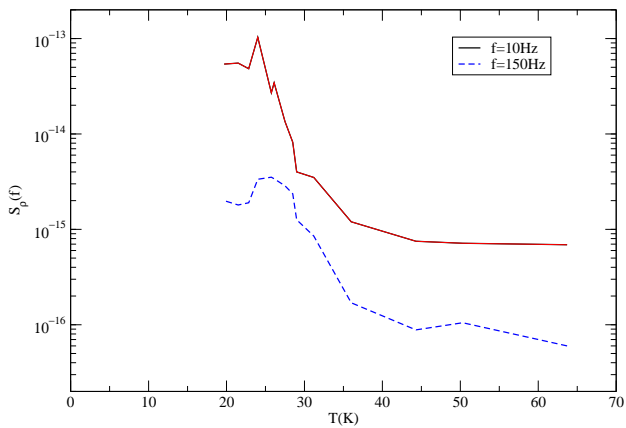}
\caption{\label{fig:temp-dep} (color online) Temperature dependence of the resistivity noise $S_{\rho}(f)$ measured at $f=10$ Hz (solid curve) and $f=150$ Hz (dashed curve).}
\end{figure}

Figure~\ref{fig:temp-dep} shows the temperature dependence of the resistivity noise at two difference frequencies. The temperature range shown in in the figure corresponds to the low frequency regime for both $f=10$ and $f=150$ Hz. This can be deduced from Fig.~\ref{fig:freq-dep} where the crossover frequency (between the low- and high-frequency regimes) for $T\sim 13$ K is around 200 Hz, and the data in Fig.~\ref{fig:temp-dep} are taken at $T > 13$ K. We analyze the temperature dependence in the context of the magnetic and charge fluctuation models we have developed above.

For models A and B as well as for the Griffiths picture, the temperature dependence of the normalized resistivity noise is directly related to the temperature dependence of the magnetic correlation length $\xi_{M},$ as well as the coefficients $\Gamma_{0}$ (Model A) and $D$ (Model B). In all cases, $\langle|\delta\rho_{\omega}|^{2}\rangle/\rho^{2}\sim\xi_{M}^{2}.$
We do not yet know the contribution to $\Gamma_{0}$ and $D$ from
the change of hole resistivity with temperature. Suppose the holes
in the quantum well do contribute to $\Gamma_{0}$ and $D.$ In this
case, as the temperature is reduced, $\Gamma_{0}$ and $D$ can show
a decrease mirroring the resistivity, and that it has an anomaly when the correlation length becomes comparable with the inter-droplet separation.

For models A and B in the low-frequency regime, in accordance with
Eqs.~\eqref{eq:r-noise2} and \eqref{eq:C-B-w-lim}, we fit the temperature dependence data in Fig.~\ref{fig:temp-dep} to $\langle|\delta\rho_{\omega}|^{2}\rangle=C\rho^{2}\xi_{M}^{2},$
where $C$ is a constant. The exponential temperature dependence of
$\rho$ and $\xi_{M}$ together with the scatter in the data makes
it difficult to extract the sub-leading power-laws and logarithms;
thus, we do not attempt to distinguish models A and B here. For the
charge fluctuation model, following Eq.~\eqref{eq:chargeflucnoise},
we tried fitting the data to $\langle|\delta\rho_{\omega}|^{2}\rangle=M\rho^{3}\exp(-2\Delta/T),$
where $M$ is a constant. We consider $T_{C}$ and $\Delta>0$ as
fitting parameters. We could not obtain a satisfactory fit for the
charge fluctuation result for any positive value of $\Delta.$

\begin{figure}
\includegraphics[width=0.95\columnwidth]{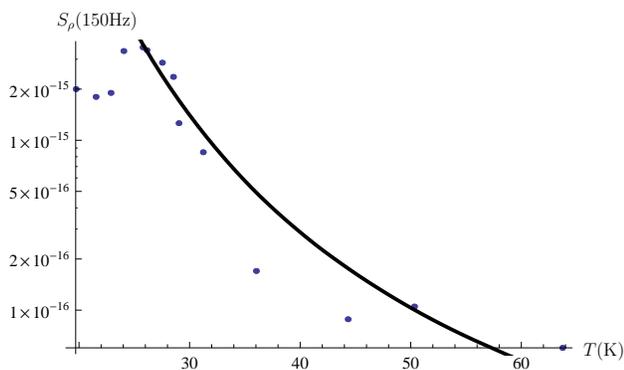}
\caption{\label{fig:tempdep-fit} (color online) Temperature dependence of the resistivity noise $S_{\rho}(f)$ for $f=150$ Hz (dots) and a fit (solid curve) to our model of magnetic fluctuations, $S_{\rho}(f)\sim C\rho^{2}e^{\pi T_{C}/T}.$ The fit corresponds to $T_{C}=53.22$ K. The model does not take into account changes in the damping parameters $\Gamma_{0}$ or $D$ across the resistivity anomaly.}
\end{figure}

Fig.~\ref{fig:tempdep-fit} shows the resistivity noise data for $f=150$ Hz and a fit to our model of magnetic fluctuations. Possible variation of $\Gamma_{0}$ or $D$ across the resistivity anomaly is not taken into account in the fit. From this fitting we have got $T_{C}=53.22$ K that by two times exceeds the temperature of the resistance anomaly and maximum of $S_{\rho}(f)$ versus temperature, while it is in a good agreement with $T_{C}=49$ K, which was found from fitting of the resistivity temperature dependence.~\cite{tripathi11} That additionally proves the earlier result that there are two characteristic temperatures in such systems: the local Curie temperature at which ferromagnetic ordering occurs inside the magnetic droplets and the global Curie temperature of the long-range ferromagnetic ordering establishment at the scale of the whole sample ``magnetic percolation transition''.

\section{Discussion \label{sec:Discussion}}

We experimentally studied and analyzed the resistivity noise in 2D DMS structures. We have shown how noise measurements can be used to probe the microscopic dynamics governing ferromagnetism in DMS heterostructures. For insulating DMS systems, we obtained a relation between the resistivity noise and the spin autocorrelation function which enables us to probe magnetization dynamics through transport measurements. We studied a number of microscopic models describing ferromagnetism in the Mn $\delta$-layer motivated by the hydrodynamic approach of Hohenberg and Halperin.\cite{hohenberg77} The models studied fall into two broad classes: (a) those where the order parameter is not a conserved quantity (Model A) such as Ising
magnets and (b) those where the order parameter is conserved (Model
B) such as uniaxial Heisenberg ferromagnets, RKKY ferromagnets, and
double-exchange ferromagnets. Model B dynamics, where observed, can
be very useful in understanding the microscopic origin of ferromagnetism. In particular, we showed that resistivity noise arising from magnetization fluctuations of RKKY and double-exchange ferromagnets are qualitatively different and can thus be used to address the long standing question of whether ferromagnetism in DMS systems arises from an RKKY or a double exchange mechanism. We also analyzed the effect of disorder on the magnetization dynamics.

Our results are summarized as follows.

(a) The resistivity noise for uniaxial Heisenberg ferromagnets as
well as disordered RKKY ferromagnets decreases as $S_{\rho}(t)\sim1/t$ for times long compared with the time scale $\tau=\xi_{M}^{2}/D$ of spin diffusion over the magnetic correlation length. In contrast, $S_{\rho}(t)\sim1/t^{2/5}$ for a double exchange ferromagnet. These cases all involve a conserved order parameter. When the order parameter is not conserved, such as in an Ising ferromagnet (or an antiferromagnet), the resistivity noise has a random-telegraph behavior and follows $S_{\rho}(t)\sim e^{-\Gamma_{0}t}$ at long times. The differences arise from the momentum dependence of the damping rate $\Gamma(\mathbf{q})$
of magnetic fluctuations. $\Gamma(\mathbf{q})=\Gamma_{0}$ for the
Ising case; $\Gamma(\mathbf{q})\sim Dq^{2}$ for uniaxial Heisenberg
ferromagnets and disordered RKKY ferromagnets; and $\Gamma(\mathbf{q})\sim Aq^{5}$ for the 2D double exchange case.

(b) Magnetic disorder in the form of site dilution and random sign
of nearest neighbor exchange interaction was considered. As long as
one remains above the classical percolation threshold, site dilution
was seen to degrade the Curie temperature and spin stiffness but otherwise retain the same dynamics as the undiluted case. This is in effect a confirmation of the validity of the Harris criterion in our systems. When the sign of inter-impurity exchange interaction varies randomly, as may be the case at low Mn density together with an RKKY interaction of the Mn atoms, the ground state is a spin glass instead of a ferromagnet. It was shown that the damping rate of magnetic fluctuations for this case also goes as $\Gamma(\mathbf{q})\sim Dq^{2}.$

(c) Fluctuations of charge in the puddles in the quantum well result
in a random telegraph noise of the form $S_{\rho}(t)\sim e^{-t/\tau}$ at long times. In terms of frequency, this is a Lorentzian behavior that has been obtained in Sec. \ref{sec:charge-fl} and discussed in relation to the experimental data in Sec. \ref{sub:freq-time-dep}.

(d) Our experimental data indicate that $S_{\rho}(t)\sim1/t$, which is evidence in favor of a disordered RKKY ferromagnet for these samples. However, as we have discussed in Sec. \ref{sec:Proximity} and Sec. \ref{sec:Models}, the RKKY behavior is not universal, and depends on sample parameters such as the ratio of the inter-hole distance and the Bohr radius in the Mn layer. The temperature dependence of resistivity noise qualitatively agrees with the theory although the agreement is not as good as the frequency or time dependence.

(e) Our method of probing the microscopic interactions governing ferromagnetism is not specific to 2D DMS structures and can prove useful in the study of other ferromagnetic systems.

\section*{Acknowledgments}

We are grateful to D. Dhar, A.M. Finkelstein, D.E. Khmelnitskii,
E.Z. Meilikhov, P. Mahadevan and T.V. Ramakrishnan for helpful discussions. The
work was supported by the Russian Foundation for Basic Research, projects Nos. 09-02-92675-IND, 11-02-00363, and 11-02-00708, and by the Indo-Russian Collaboration Program, grant No. INT/RFBR/P-49. We also acknowledge the partial support of EuroMagNET under EU Contract No. RII3-CT-2004-506239. V.T. and K.D. acknowledge the support of TIFR.

\end{document}